# Effect of Cd diffusion on the electrical properties of the Cu(In,Ga)Se$_2$ thin-film solar cell

Anna Koprek[a], Pawel Zabierowski[b], Marek Pawlowski[b], Luv Sharma[e], Christoph Freysoldt[a], Baptiste Gault[a,d], Roland Wuerz[c], Oana Cojocaru-Mirédin[a,f,*]

[a] Max-Planck-Institut für Eisenforschung GmbH, Max-Planck-Straße 1, 40237, Düsseldorf, Germany

[b] Faculty of Physics, Warsaw University of Technology, Koszykowa 75, 00-662 Warsaw, Poland

[c] Zentrum für Sonnenenergie- und Wasserstoff-Forschung Baden-Württemberg, Meitnerstrasse 1, 70563 Stuttgart, Germany

[d] Department of Materials, Royal School of Mines, Imperial College, Prince Consort Road, London, SW7 2BP, UK

[e] Department of Mechanical Engineering, Technische Universiteit Eindhoven

[f] I. Institute of Physics, RWTH University Aachen, Sommerfeldstraße 14, 52074 Aachen, Germany

**Corresponding Authors:** cojocaru-miredin@physik.rwth-aachen.de;

**Abstract**

Cu(In,Ga)Se$_2$ (CIGSe)-based solar cells are promising candidates for efficient sunlight harvesting. However, their complex composition and microstructure can change under operation conditions, for instance heating from sun light illumination can lead to a degradation in performance. Here, we investigate the thermally-induced degradation processes in a set of CIGSe-based solar cells that were annealed at temperatures between 150°C and 300°C. Using correlative atom probe tomography (APT)/transmission electron microscope (TEM), we found that the buffer/absorber interface is not sharp but consists of an interfacial zone (2 - 6.5 nm wide) where a gradient of constituent elements belonging to the CdS buffer and CIGSe absorber appears. An enhanced short-range Cd in-diffusion inside the CIGSe was observed whenever a low Ga/(Ga+In) ratio (≤ 0.15) occurred at the interface. This might indicate the presence of Ga vacancies as a channeling defect for Cd in-diffusion inside the CIGSe layer leading to a buried p/n-homojunction. We evidence that a considerable amount of Cd is found inside the CIGSe layer at annealing temperatures higher than 150°C. Further investigations of the elemental redistribution inside the CIGSe layer combined with C-V measurements support the formation of Cd$_{Cu}$$^+$ donor like defects deep inside the *p*-type CIGSe which lead to a strong compensation




of the CIGSe layer and hence to strong deterioration of cell efficiency at annealing temperatures higher than 200°C. Hence, understanding the degradation processes in Cu(In,Ga)Se$_2$ (CIGSe)-based solar cells opens new opportunities for further improvement of the long-term device performance.

# 1 Introduction

Further improvement in the solar cell technologies strongly rely on understanding the mechanism for the cell degradation induced by environmental conditions such as temperature, humidity, as well as the build-up of electrostatic or chemical potential during operation. Yet, to identify the primary cause for the solar cell degradation, detailed characterization at the nanometer level must be deployed to reveal, for example, the inter-diffusion phenomenon taking place between different layers inside the cell.

Cu(In,Ga)Se$_2$ (CIGSe)-based solar cells are amongst the highest performing thin-film solar cells. Besides their intrinsic high efficiency (up to 23.35 % [1]), the device is characterized by high absorption coefficient of the absorber layer ($> 10^5 cm^{-1}$[2]), long term stability in the range of 10–15 years [3] and low cost of production (0.4 US $ / W [4]). Further improvement of the efficiency of CIGSe solar cells resides mainly in the control of the chemical composition across the heterojunction.

Although a lot of care has been taken in the past when performing compositional characterizations of the CdS/CIGSe interface, it is well known that this interface is very rough, with root mean square roughness between 7 and 75 nm [5, 6], hindering precise chemical analysis. For example, the composition of the CdS/CIGSe interface determined using standard characterization techniques, such as x-ray photoelectron spectroscopy (XPS) [7, 8] or energy dispersive X-ray (EDX) microanalysis [9-11], is very often biased due to the overlap between the buffer and absorber layer (i.e. limited spatial resolution). Atom probe tomography (APT) can overcome this limitation [12-16] and map the interface in three-dimensions [17], in particular correlated with transmission electron microcopy (TEM)[18-26] or other microscopy technique [27-33] to understand the relationship between composition and material property. APT studies of the CdS/CIGSe interface have been already conducted in the past [34-37]. Here now the impact of Cd diffusion (measured by APT) on cell efficiency will be investigated and analyzed in more detail.

Therefore, here, we focus on the thermally-induced degradation processes occurring across the CdS/CIGSe interface. We report on a correlative APT/TEM analysis of the CdS/CIGSe interface morphology. We introduce an analytical method specifically developed to improve the compositional measurements by alleviating the effects of peaks overlaps in the APT mass spectrum, i.e. ions from different species but with the same mass-to-charge ratio. We propose a model for Cd in-diffusion inside the CIGSe absorber layer based on our experimental results. The APT profiles are discussed in terms of effects occurring at different length scales. Temperature induced changes of opto-electrical properties and charge carrier doping are further discussed in terms of observed chemical changes. Based on these results, we propose an explanation for thermally-induced degradation processes and an alternative perspective for further development of CIGSe-based solar cells.



## 2 Material and methods
### 2.1 Sample preparation

The CIGSe solar cells were manufactured at the Zentrum für Sonnenenergie- und Wasserstoff-Forschung Baden Württemberg in Stuttgart, Germany. The stack ZnO(Al)/i-ZnO/CdS/CIGSe was deposited on the top of a Mo coated soda lime glass substrate. The CIGSe absorber layer was grown using a multistage inline co-evaporation process at approx. 600°C following the procedure described in Ref. [38]. The resulting thickness of the CIGSe layers measured by X ray fluorescence spectrometry was 2.4µm and their composition was Cu/(In+Ga) = 0.86 and Ga/(Ga+In) = 0.29. CdS was deposited using chemical bath deposition at 65°C, whereas intrinsic ZnO and ZnO:Al layers were deposited by RF and DC sputtering. Finally, heat treatments were conducted. Hence, the samples obtained were regrouped into five sample sets: one non-annealed named here "Reference" and four annealed for 30 minutes in the dark at temperatures of 150°C, 200°C, 250°C in air and 300°C in vacuum. The cells which were annealed are named here Annealed at 150°C, Annealed at 200°C, Annealed at 250°C, and Annealed at 300°C.

To facilitate the site-specific specimen preparation for APT, the ZnO layers were selectively etched in diluted HCl (1%), and four additional CdS layers were deposited serving as protection against environmental conditions and simplifying the FIB preparation of APT tips including the Cd/CIGS interface. Additionally, 210 nm of Cr were sputter deposited as a capping layer protecting against the damages induced by the energetic $Ga^+$-beam during FIB milling. To avoid sample heating, Cr was deposited in three steps separated by a 20 min break. One needs to mention here that the surplus in CdS and the Cr layer were deposited after conducting the heat treatments so that the higher amount of Cd will not be artificially inserted in the absorber layer.

The cells intended for electrical characterizations were completed with the Ni/Al/Ni contact grids deposited by electron beam deposition and mechanically structured leading to a total cell area of 0.5 $cm^2$, as schematically depicted in Figure 1.

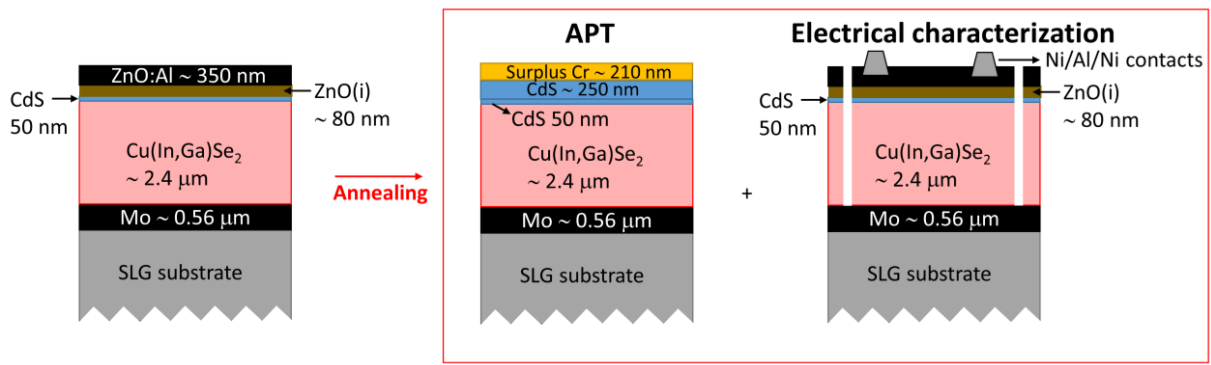

**Figure 1: Description of the CIGSe solar cell and of samples used for APT and electrical characterization.** (Left): the structure of the CIGSe solar cell with the stacking Mo-coated SLG (soda lime glass)/Mo (0.56 µm)/Cu(In,Ga)Se$_2$ (2.4 µm)/CdS(50 nm)/i-ZnO (80 nm)/ZnO:Al (350 nm). (Right): the schematics of samples aimed for chemical and electrical investigations. In total five samples sets were investigated: one non-annealed called "Reference" and four others annealed at 150°C, 200°C, 250°C, and 300°C. A special sample preparation was fulfilled



for the APT characterization after performing the heat treatments: the ZnO layers were selectively etched in diluted HCl (1%), and four additional CdS layers were deposited in order to simplify the FIB preparation of APT tips including the Cd/CIGSe interface. The samples intended for the electrical characterization were simply completed with the Ni/Al/Ni contact grids deposited by electron beam deposition and mechanically structured leading to a total cell area of 0.5 cm$^2$.

## 2.2 Electrical characterizations.

The temperature dependent current density-voltage (*J(V,T)*) characteristics were measured both in dark and under standard illumination condition (integrated light power of 100 mW·cm$^2$) at temperatures of 288 K, 293 K, 303 K, 313 K, 323 K, 333 K and 341 K. Owing to the metastable behaviour of CIGSe-based solar cells with respect to the applied illumination and bias [27], all dark current density-voltage (*J(V)*) curves were measured first, i.e. before the illuminated *J(V)* and external quantum efficiencies (EQE). The dark temperature dependent J(V,T) characteristics were used to determine the ideality factor (A) of each solar cell using the methodology described in Ref. [39]. Similarly, the illuminated J(V) curves were used to determine the activation energy ($E_a$) for the recombination mechanism using the formalism of Shockley and Queisser detailed in Ref. [40]. The obtained values for the activation energy ($E_a$) are given in Table 1.

Current density voltage (JV) characteristics of the solar cells were measured under illumination with simulated AM 1.5 spectrum at standard test conditions. The EQE studies were performed with the BENTHAM PVE 300 system supported with Keithley 2400 Source Meter. For Admittance Measurements (AS) the capacitance (C) was measured using an Agilent E4980A LCR meter as a function of frequency (f) and temperature (T). The width of the space charge layer (W) was evaluated using the depletion approximation according to the formula W = $\varepsilon_0\varepsilon_r$S/C, where $\varepsilon_r$ is the absorber dielectric constant, $\varepsilon_0$ is the vacuum permittivity, and S is the sample area.

## 2.3 Atom probe tomography – data acquisition and reconstruction

All APT samples were prepared using a dual focused ion beam (SEM/FIB) Helios Nano Lab 600i following the lift-out method described in Ref. [41]. By means of annular milling, sharp needle-shape specimens were formed with an apex diameter < 100 nm. Subsequent steps of specimen thinning were done under reduced acceleration voltage of 16 kV, since the CdS layer is particularly sensitive to Ga$^+$ beam milling. In the end low-energy (5 keV) cleaning was performed in order to minimize material damages induced under higher energy beam (16 keV) conditions. APT experiments were performed on a CAMECA LEAP 3000X HR instrument equipped with a pulsed laser source, with a pulse duration of 12 ps, at a wavelength of 532 nm. To limit the influence of the preferential field evaporation of one or more of the constituent elements, in particular in between the laser pulses, calibrated parameters were used providing accurate stoichiometry of the investigated CIS material [42]. These parameters are: 0.1 nJ laser pulse energy, 100 kHz pulse repetition rate, and 60 K base temperature. The cryogenic temperature was kept constant throughout the experiment to minimize possible surface diffusion of the constituents induced by laser pulses or the electrostatic field. To ensure reliable statistics, at least three successful APT measurements were performed for each sample.



So as to maximize the reliability of the APT data reconstruction, the parameters were calibrated by performing correlative APT-TEM investigations on three different APT specimens [43]. For this purpose, a 2D projection of the CdS/CIGSe interface was obtained by TEM using a JEOL 2200 FS at 200 kV. A set of 3D APT reconstructions was then created using the commercial software IVAS 3.6.10 provided by Cameca Instruments Inc. with systematic variation of the reconstruction parameters, i.e. image compression factor ICF and field reduction factor k (see Fig. S1 and Fig. S2 in supplementary information for details). The best match between the reconstructed APT data and the acquired TEM micrograph of the CdS/CIGSe interface (see Fig. S2 in supplementary information) reveals the ICF and k parameters that were used for further APT investigation, i.e. ICF= 1.4 and *k* = 4 (Fig. S2).

*Atom probe tomography – new data processing routine*

A typical issue with APT measurement is the overlap between peaks in the mass spectrum due to isobars or combination of elements in molecular ions that have a mass-to-charge ratio similar to that of another element. We developed a method that uses the natural abundancies of the different elements in the measured APT dataset to recalculate the local composition within each of the bins of a composition profile. Here, typical overlaps are encountered for: Cu and S at 32.5 Da (Da stands for Dalton) and 65 Da as well as Cd and In at 56.5 and 113 Da. These overlaps make the association of a given peak to a specific atomic or molecular species challenging. Developing and applying this method is necessary to obtain more accurate compositional values. We refer here to peak deoverlapping since no true deconvolution procedure, in the mathematical sense, is used.

Here we developed a new "peak deoverlapping" protocol (for details see supplementary materials, Fig. S3 to Fig. S7) which subtracts the background and estimates the precise amount of each element observed in the mass spectrum. Fig. 2 shows an example of 1D composition profile before and after applying the peak deoverlapping method. It can be seen that the elemental composition profiles in the CdS layer are slightly modified giving reliable results for the composition of the CdS layer and for the Cd composition.

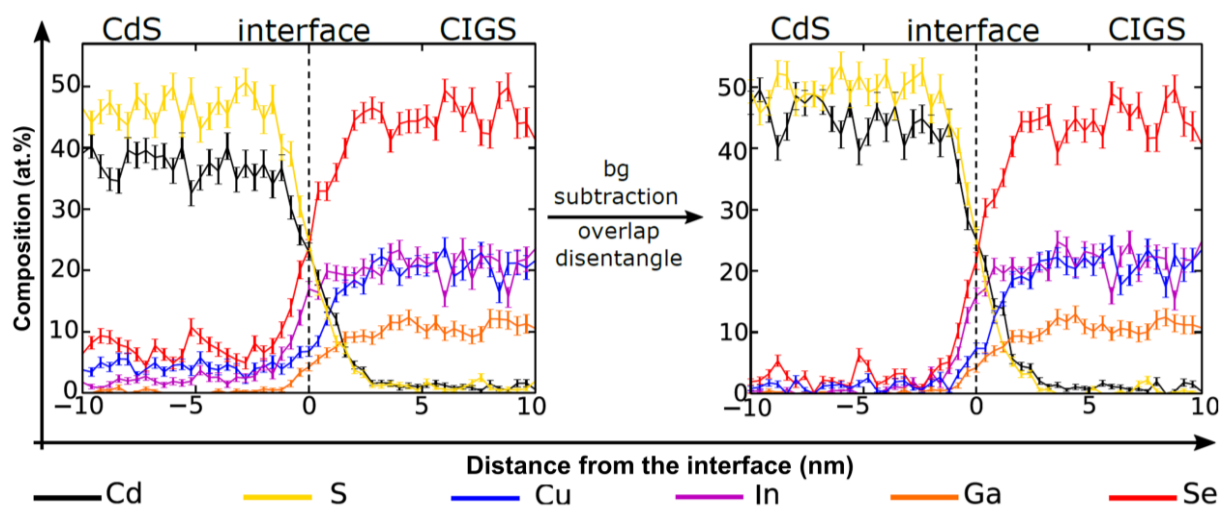



**Figure 2: 1D composition profiles across a flat region of the CdS/CIGSe interface before and after background subtraction and deoverlapping.** This method clearly shows the reduction of Cu, In and Se signal at the CdS side and, hence, the improved stoichiometry of the CdS phase.

## 3 Results and Discussion

### 3.1 Electrical characterization

To characterize the opto-electrical parameters and transport mechanism of the investigated solar cells, a series of electrical measurements were performed on each of the heat treated samples. Figure 3 reports: i) illuminated current density-voltage curve (J(V)) measured at room temperature in (a); ii) open circuit voltages ($V_{oc}$) as function of the temperature (between 288 K and 341 K) in (b); iii) external quantum efficiency (EQE) in (c).

The illuminated J(V) curves in Fig. 3a. revealed progressive degradation of the cell efficiency, i.e. a decrease in the $V_{oc}$ and fill factor (FF), with increasing annealing temperature. A strong degradation of the cells is particularly visible for annealing temperatures of 250 °C and 300 °C for which the drop in $V_{oc}$, compared to the reference sample, amounts to 0.18 V and 0.30 V, respectively (see Table 1). Simultaneously the value of the short circuit current density ($J_{sc}$) changes negligibly. Yet, the EQE characteristics (Fig. 3c.) barely changes with the applied heat treatment showing very little difference in the probability of light absorption and carrier collection at the back-side of the device. Only a small decrease (of approx. 2 %) in the EQE signal was observed for the 300°C sample at the wavelength of incident light between 550 nm and 650 nm. Towards the long-wavelength limit, the EQE curves are almost identical, which indicates the absorber bandgap in all five samples. When plotting $(E*EQE)^2$ over E, where E is the energy of the incoming light, and extrapolating the curve towards the energy axis (not shown here), we determined the value of the absorber bandgap $E_g$ to be equal to 1.15 eV [40].

To find out more about the dominant recombination mechanisms that takes place inside each sample, two additional parameters were determined from illuminated and dark J(V,T) measurements: the activation energy for recombination $E_a$ and the ideality factor A of the solar cells. The $E_a$ indicates the location of recombination process inside the solar cell. More precisely, bulk recombination is the most dominant recombination process when $E_a$ is closer to the band gap value of 1.15 eV. If $E_a$ is substantially smaller than the bandgap value then interface recombination must be the most dominant recombination mechanism [44]. The activation energies for the samples annealed at temperatures lower than 250°C are 0.15eV higher than the bandgap of 1.15eV extracted from the EQE measurements. This is caused by the fact, that $V_{OC}$ is not only exponentially depending on temperature via the activation energy, but is also depending on saturation current density which is proportional to $T^{3/2}$ leading to this small gap between $E_g$ and $E_a$. The ideality factor A indicates the character of the predominant recombination mechanism: i) radiative and interface for A = 1, ii) Shockley Read-Hall in the space charge region for $1 \leq A \leq 2$, and iii) Auger or tunneling enhanced recombination for $A \geq 2$. Except for the sample annealed at 300° C the values of ideality factors in the whole temperature range are between 1.2 and 1.4 indicating that Shockley-Read-Hall (SRH) recombination dominates. Only for the sample annealed at 300°C the ideality factor exhibits stronger temperature dependence with a value exceeding 2 which points out at a tunneling assisted recombination process.



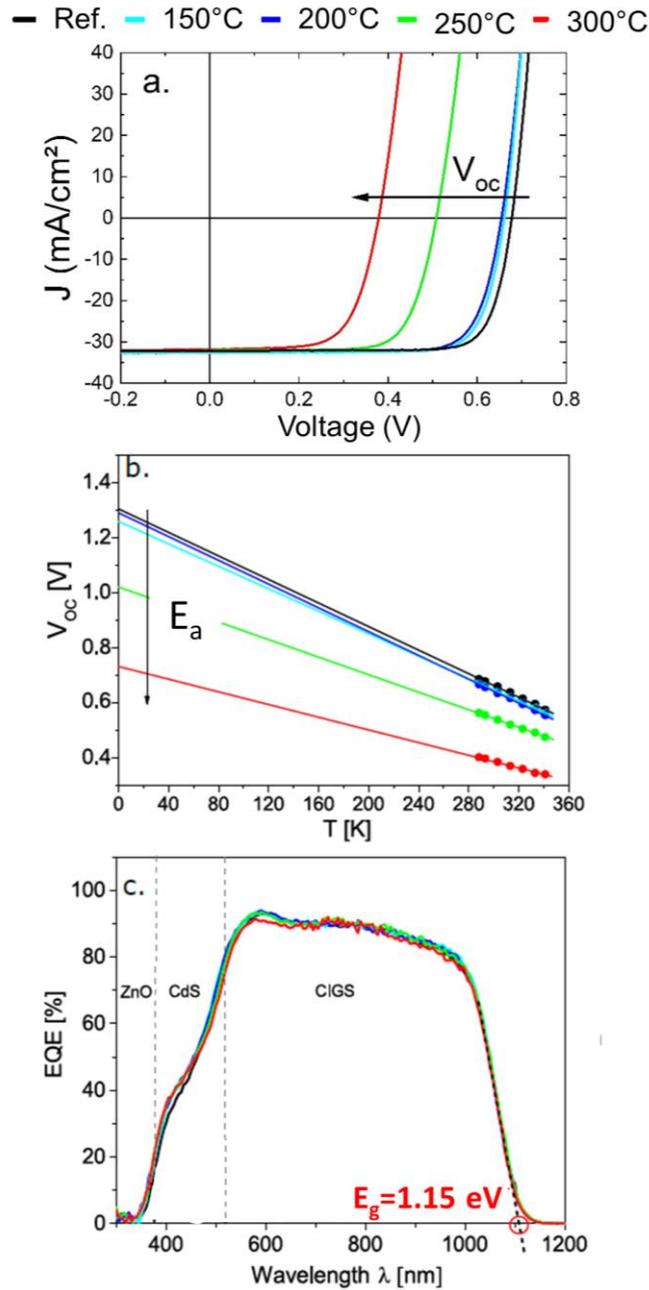

**Figure 3: Electrical characterization of the reference cell as well of the cells annealed at temperatures between 150°C and 300 °C.** a. J-V characteristics of the best cell of each sample as measured at room temperature showing the decrease in open circuit voltage ($V_{oc}$) and fill factor (FF) with increasing annealing temperature; (b.) $V_{oc}$ as a function of temperature showing the change in recombination mechanism with the applied annealing temperature; (c.) room-temperature external quantum efficiency (EQE) showing almost no change in charge carrier collection, the two vertical dashed lines mark the position of the bandgap $E_g$ of ZnO and CdS.

Table 1 shows that $E_a$ decreases from ~1.3 eV to 0.76 eV when the annealing temperature increases beyond 200°C. This strong drop of $E_a$ is a signature for the change in recombination pathway from bulk at Reference, 150°C and 200°C, throughout intermediate (bulk and



interface) at 250°C, to interface recombination at 300°C. Hence, these electrical characterizations proved indeed that the cell efficiency drops (cell degradation) along with the annealing temperature, and is correlated with a change in the predominant recombination mechanism from bulk to interface recombination. To find out the mechanism responsible for the degradation of the cell performance with the annealing temperature, APT and AS are used to track also the thermally induced changes in chemical composition and carrier concentration inside the solar cells.

**Table 1:** Solar cell parameters of CIGSe solar cells annealed for 30 min at different temperatures. Efficiency $\eta$, open circuit voltage $V_{oc}$, Fill Factor *FF* and short circuit current density $J_{sc}$, mean values of up to ten cells per samples (shunted cells were excluded). Room-temperature hole concentration $p$ after 30 min of light soaking under AM1.5 conditions and extrapolated activation energy $E_a=V_{oc}$ $(T=0K)$ as well as the average Cd composition [Cd] inside the CIGSe layer obtained by APT are also provided.

| Sample | $\eta$ (%) | $V_{oc}$ (mV) | FF (%) | Jsc (mA/cm²) | p (cm$^{-3}$) | $E_a$ (eV) | [Cd] (at.%) |
|---|---|---|---|---|---|---|---|
| Reference | 17.2 | 673 | 80.0 | 32.0 | 6.2x10$^{15}$ | 1.29 | 0.21±0.04 |
| Annealed at 150°C | 16.8 | 661 | 78.5 | 32.4 | 2.5x10$^{15}$ | 1.24 | 0.35±0.06 |
| Annealed at 200°C | 16.3 | 649 | 77.2 | 32.5 | 2.0x10$^{15}$ | 1.26 | 0.5±0.08 |
| Annealed at 250°C | 11.4 | 492 | 70.8 | 32.6 | 1.7x10$^{15}$ | 1.04 | 0.66±0.05 |
| Annealed at 300°C | 7.4 | 372 | 63.9 | 31.3 | 1.3x10$^{14}$ | 0.76 | 0.99±0.14 |

### 3.2 Interface chemistry

Fig. 4 shows a selection of three 1D composition profiles across a flat region of CdS/CIGSe interface obtained from different APT measurements for the reference sample and the samples annealed at 200°C and 300°C. These profiles are calculated using the proximity histogram approach [45], i.e. as a function of an iso-composition surface of 25 at.% Cd. The green band highlights the interfacial region where either inter-diffusion or depletion of one of the constituent elements occurs. As presented in Fig. 4, the width of the interfacial band differs from one profile to another and does not depend on the applied heat treatment but rather on the region of the solar cell from which the APT specimen was prepared. Across all composition profiles, the width of the interface band spans from 2 nm to 6.5 nm, with no apparent correlation with the annealing temperature. Surprisingly, the heat treatment did not influence the width of the interfacial zone even at a relatively high temperature of 300°C. A certain width of the interface of approximately 1.5 nm is expected at each of the annealing temperatures due to surface reconstruction that is attributed to the presence of dipole momentum coming from either cation (Cu or In/Ga) or anion (Se) terminated {112} surfaces [46, 47]. The relaxation of the surface (interface) dipole moments often occurs via chemical intermixing of the first few atomic



planes between both sides of the interface. Most likely this effect can also be observed in the APT composition profiles between -1 to 1 nm interval around the interface position (0 nm) where CdS constituents intermix with CIGSe constituents. Yet, the total width of the interfacial zones is much larger than the expected width due to intermixing. In addition, in most of the composition profiles, the interface is asymmetric being much wider on the CIGSe side of the interface. Moreover, the width of the interface region does not only change from one sample to another, but also within the same sample (from one region of the sample to another region).

Nevertheless, certain similarities were observed in the profiles shown in Fig. 4. The most important similarity observed in almost all profiles is the coexistence of a 1–4 nm wide region, in which both Cu and Ga are depleted at the CIGSe surface (in most of the profiles from Figure 4, except for Fig. 4 d. and Fig 4.f). Another important similarity is the in-diffusion of Cd inside the CIGSe layer observed interestingly for the same profiles where Cu and Ga were depleted indicating thus a clear correlation between Cd and Cu/Ga. In Fig. 4.d and 4.f the Cd diffusion inside the CIGSe layer is only within 1-2 nm, which is in the same range of order with the layers intermixing and, hence, no clear indications regarding Cd diffusion can be made.

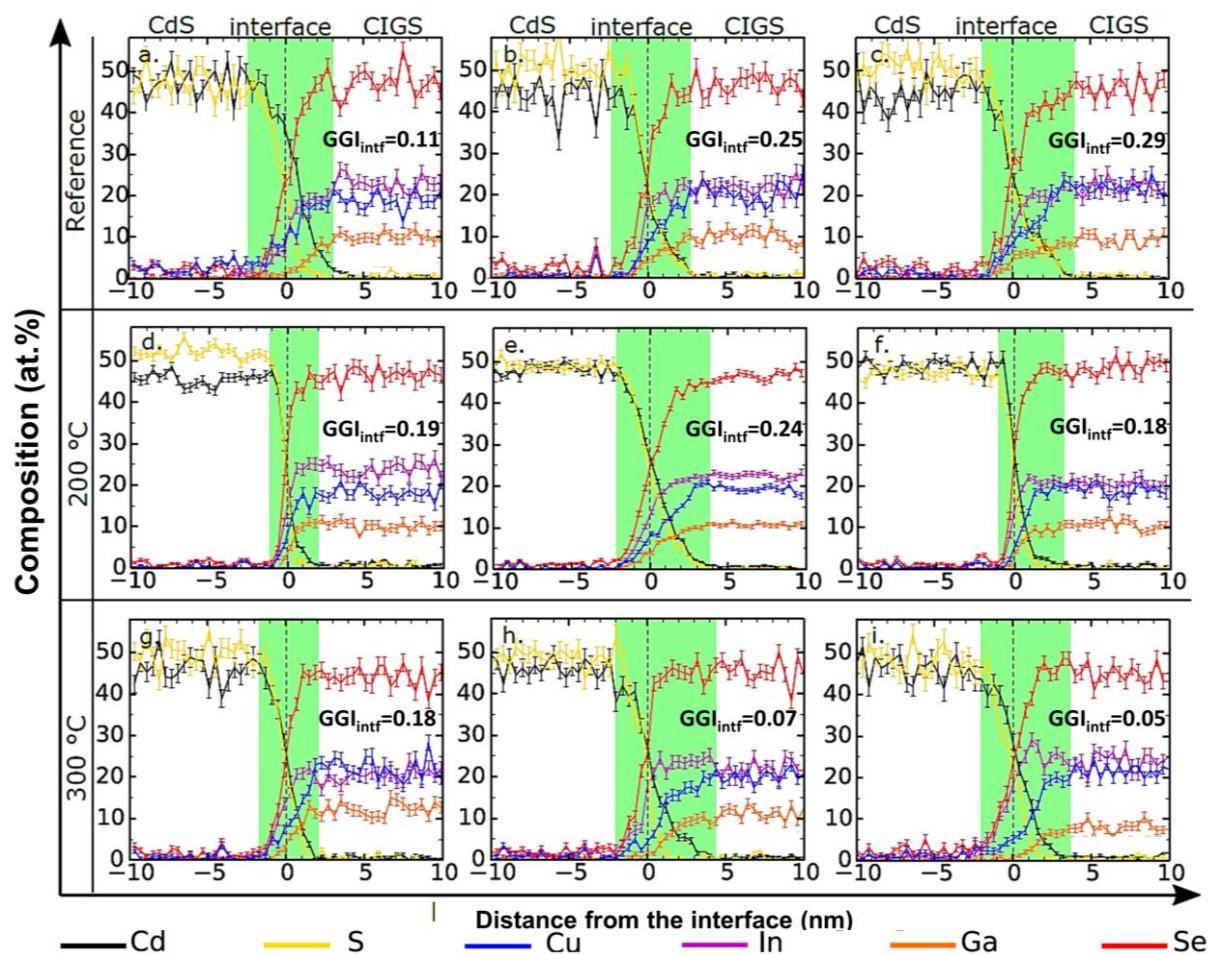

**Figure 4: 1D composition profiles derived from APT measurements**. These profiles were constructed across the flat region of the CdS/CIGSe interface for the reference sample and the samples annealed at 200°C and 300°C. The Green band represents the interface region for which depletion or in-diffusion of one of the constituent elements was identified.



Corresponding Ga/(Ga+In) ratio calculated at the absorber surface and abbreviated by GGI$_{intf}$ is given for each proxigram.

Comparative analysis of all composition profiles revealed that Cd in-diffusion into the near surface area of CIGSe takes place when low values ($\leq 0.3$) of Ga/(Ga+In) ratio (abbreviated GGI$_{intf}$ where "intf" stands for interface in Figure 4) are detected at the CdS/CIGSe interface. The literature data included in table 2 (both experimental and theoretical) consistently show that the Ga content inside the CIGSe influences mainly the position of the conduction band edge, whereas the Cu content influences the position of the valence band edge at the CdS/CIGSe interface [48, 49]. For a Ga content of x= 0.2 (GGI=Ga/(Ga+In)=0.2), the CdS/CIGSe band alignment forms a spike-like configuration with positive conduction band offset (CBO) (determined from CBO = CBM$_{CdS}$ - CBM$_{CIGSe}$) of +0.3 eV [48, 49]. Further increase of Ga composition causes an upward shift of the CIGSe conduction band edge leading first to well aligned bands and further to a cliff-like configuration, with negative CBO of -0.2 eV for a Ga content of x= 0.75 [48, 49]. On the contrary, the Cu content influences the position of the CIGSe valence band edge. In a stoichiometric CIGSe material the Se *p* and Cu *d* orbitals form the valence band maximum via so called *p-d* coupling. The interaction between Se *p* and Cu *d* orbitals causes an upward shift of the valence band edge as compared to the isolated Se *p* and Cu *d* states [47, 50]. Therefore, the calculated valence band offsets (VBOs) between CdS and CISe (stoichiometric compounds) [51, 52] are higher than the experimentally observed values of ∼ 0.8 eV, that were mostly reported for Cu depleted surfaces [47, 50]. Yet the depletion of Cu will break the *p-d* interaction (due to absence of Cu *d* orbitals), which results in a downward shift of the valance band maximum.

The observed depletion of Ga and Cu atoms in the vicinity of CdS/CIGSe interface as observed in Figure 4 will lower both the conduction band minimum (CBM) and valence band maximum (VBM), and hence leads to a CBO and VBO. Moreover, the detected weak depletion of Ga of $x \leq 0.3$ will lead to either well aligned or spike configuration between CdS and CIGSe conduction bands, which may hinder electron transport depending of the magnitude of the CBO (for low CBO of +0.3 eV [48, 49] electron transport is still possible by tunneling and/or thermionic emission). But this possible drawback is more than compensated by the reduced recombination at the CdS/CIGSe interface due to the increased type inversion at the absorber surface. More precisely, Cd in diffusion within the first few nanometers of the absorber layer can lead to the formation of Cd$_{Cu}^{+}$ donor like defects and, hence, to a reduced net doping or even type inversion near the CdS/CIGSe interface [53].

**Table 2: Literature data showing influence of Ga content on band alignment between CdS and CdS=Cu(In$_{1-x}$,Ga$_x$)Se$_2$ as well as between CuInSe$_2$ and CuGaSe$_2$.** Despite different fabrication methods and environmental conditions, the data show elevation of the CBM$_{CIGSe}$ with increasing Ga content and small effect of Ga addition on the position of VBM$_{CIGSe}$ in the vicinity of CdS phase.

| Sample | CBO (eV) | VBO (eV) | Fabrication process | Investigation method | Reference |
|---|---|---|---|---|---|
| **CdS/Cu0:93(In0:80,Ga0:20)Se$_2$** | 0.3 | 0.7 | CBD/3-stage coevaporation | PES and IPES | [48, 49] |



| | | | | | |
|---|---|---|---|---|---|
| **CdS/Cu0:93(In60,Ga40)Se$_2$** | 0-0.1 | 0.9 | CBD/3-stage coevaporation | PES and IPES | [48, 49] |
| **CdS/Cu0:93(In0:25,Ga0:75)Se$_2$** | -0.2 | 0.9 | CBD/3-stage coevaporation | PES and IPES | [48, 49] |
| **CdS/CuGaSe$_2$** | - | 0.98±0.1 | Cu-poor surface | - | [51] |
| **CdS/Cu(In0:78,Ga0:28)Se$_2$** | - | 0.88±0.1 | | - | [51] |
| **CdS/CuInSe$_2$ (001)** | 0.31±0.25 | 0.79±0.15 | PVD/epitaxial growth (Cu-poor surface) | | [52] |
| **CdS/CuGaSe$_2$** | -0.19±0.1 | 0.93±0.1 | | | [54] |
| **CdS/CuInSe$_2$** | - | 1.07 | stoichiometric | DFT-LDA | [55] |
| **CdS/CuInSe$_2$** | - | 1.13 | stoichiometric | | [56] |
| **CdS/CuIn$_5$Se$_8$** | - | 0.65 | - | DFT-LDA | [55] |
| **CuInSe$_2$/CuGaSe$_2$** | 0.6 | 0.04 | | DFT-LDA | [55, 57] |
| **CuInSe$_2$/CuGaSe$_2$** | 0.53 | 0.05 | | DFT-HSE06, CNL alignment method | [56] |

Interestingly, none of the 1D composition profiles presented in Fig. 4 revealed presence of other phases at the CIGSe side of the interface, such as CuIn$_3$Se$_5$ or CuIn$_5$Se$_8$ ordered defect compounds as was observed when employing In$_2$S$_3$ buffer [58, 59]. This is in agreement with our previous works [34-36]. Instead a gradient in composition of the constituent elements was consistently observed. The Ga and Cu depletion explains well the low values of CBO and VBM reported by other authors [51, 52], which is however not associated with the formation of ordered defect compound, but rather an intermediate zone in which electronic structure of both CdS and CIGSe materials can adjust with each other.

### 3.3 Long range Cd composition profiles

The most pronounced feature in the long range 1D composition profiles from APT is the in-diffusion of Cd inside the CIGSe. The Cd composition in CIGSe progressively increases with the annealing temperature and goes along with the degradation of the cell performance (Table 1). For the reference sample and the sample annealed at 150°C, the Cd profile decays to 0 at. % right after crossing the interface (not shown here). After annealing at 200°C, Cd diffuses into the CIGSe layer up to maximum 100 nm (as shown in Fig. 5). For samples annealed at 250°C (Fig. 5b), the Cd diffuses further decaying at ∼ 200 nm inside the CIGSe, whereas at 300°C (Fig. 5c.) the Cd composition does not decay at all to 0 at. % within the distance ranges measured with APT (maximum 500 nm deep inside the CIGSe). This is in agreement with the work of Ishizuka et al. [11] where the cell was heat-treated at 500°C. By using EDX and electron beam induced current (EBIC), the authors have found an excessive diffusion of Cd inside the absorber layer which leads to cell degradation.



Interestingly the shape of Cd in-diffusion profiles changes from "Fick's law" like to "hill" like inside the CIGSe for samples that were annealed at elevated temperatures (250°C and 300°C). This could be explained by the fact, that the supply of Cd was not sufficient enough at these higher temperatures to act as a constant source. Similar "hill" like diffusion profiles were observed for the diffusion of Na in single crystals [60].

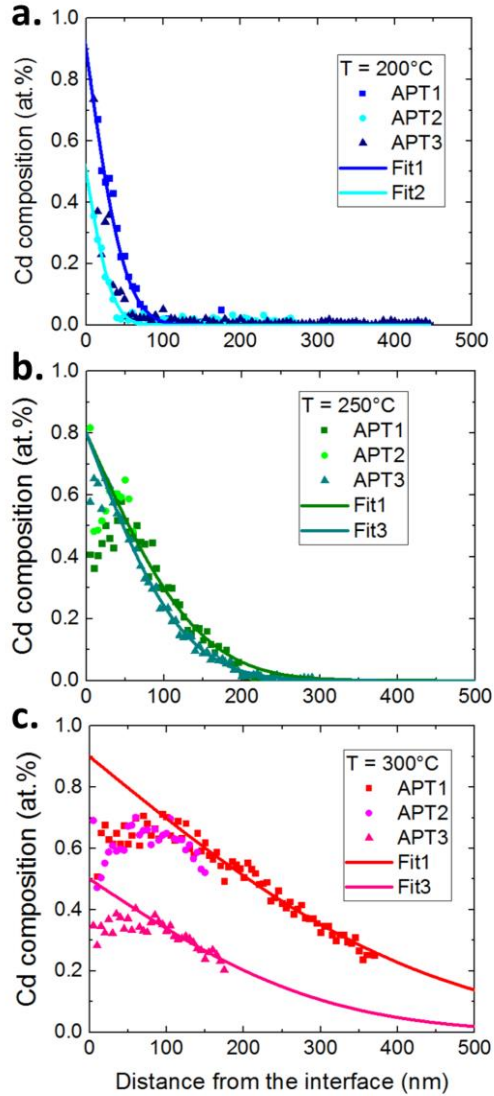

**Figure 5: Cd diffusion profiles inside CIGSe from APT measurements.** 1D composition profiles of Cd inside the CIGSe obtained for three different regions of the same sample measured by APT (symbols) and fit with equation (1) (solid lines) for the samples annealed at 200°C in (a.), 250°C in (b.), and 300°C in (c.).

These Cd in-diffusion profiles obtained by APT were fitted using a complementary error function (the solution of Fick's law for a constant diffusion source [61]):

$$C(x,t) = C_0 \, \text{erfc}\left(\frac{x}{2\sqrt{D_V t}}\right) \tag{1}$$



where $x$ describes the Cd penetration depth, $t$ is the annealing time, $D_V$ is the Cd diffusion coefficient in the CIGSe volume, whereas $C_0$ is the average Cd composition at the CIGSe surface. The diffusion coefficient extracted from the fit of the APT profiles (open triangles in Fig. 6) agree with values reported by Hiepko et al. [62] (solid line in Fig. 6 and values in table 3). Hence, this supports that APT can describe precisely the Cd volume diffusion inside the CIGSe absorber.

**Table 3**: Cd diffusion coefficients in CIGSe extracted from the Cd diffusion profiles of APT (Fig. S8) according to equation (1) with volume diffusion coefficient $D_V$ and diffusion coefficient $D_2$ (probably GB diffusion). For comparison also the literature data for $D_V$ from Hiepko et al. [62] are given.

| Sample | $D_V$ (APT) (cm²/s) | $D_V$(Hiepko) (cm²/s) |
|---|---|---|
| **Annealed at 200°C** | 4.4·10⁻¹⁵ | 4.0·10⁻¹⁵ |
| | 2.0·10⁻¹⁵ | |
| **Annealed at 250°C** | 3.5·10⁻¹⁴ | 4.6·10⁻¹⁴ |
| | 2.6·10⁻¹⁴ | |
| **Annealed at 300°C** | 3.4·10⁻¹³ | 3.4·10⁻¹³ |
| | 1.6·10⁻¹³ | |

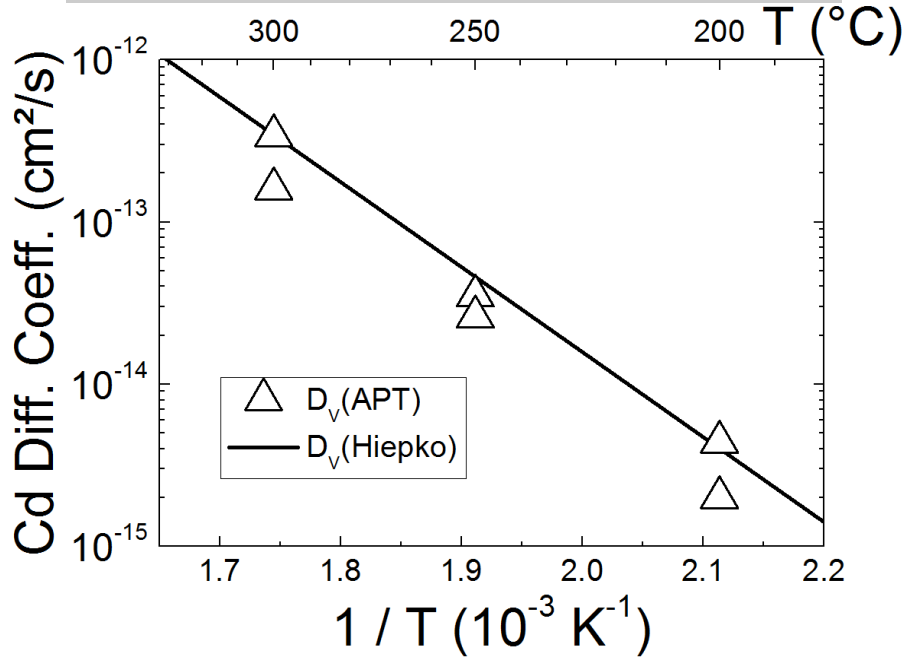

**Figure 6: The volume diffusion coefficient $D_V$ (triangles) of Cd in CIGSe plotted against the inverse diffusion temperature 1/T.** The diffusion coefficients $D_V$ shows an Arrhenius-type temperature dependence. The black open triangles are data points from APT measurements from fit of the curves in Fig. 5. The solid black line shows the diffusivity of Cd in the volume according to Ref. [62].



## 3.4 Space charge region width and open circuit voltage vs. charge carrier concentration

Fig. 7 presents the space charge region (SCR) width at room temperature derived from CV measurements at room temperature as function of the annealing temperature. The width of the SCR, i.e. depleted region in charge carriers, increases with the annealing temperature up to the complete CIGS layer thickness of 2.4 µm for the sample annealed at 300°C. Table 1 presents the free hole concentrations $p$ measured by capacitance profiling at room temperature after 30 min of light soaking under AM1.5 conditions. We note here that following the arguments presented in [63] we used values of $p$ determined after light soaking because we wanted to ensure that the free carrier concentrations are as close as possible to the net acceptor concentrations established under open circuit conditions at 300 K under illumination.

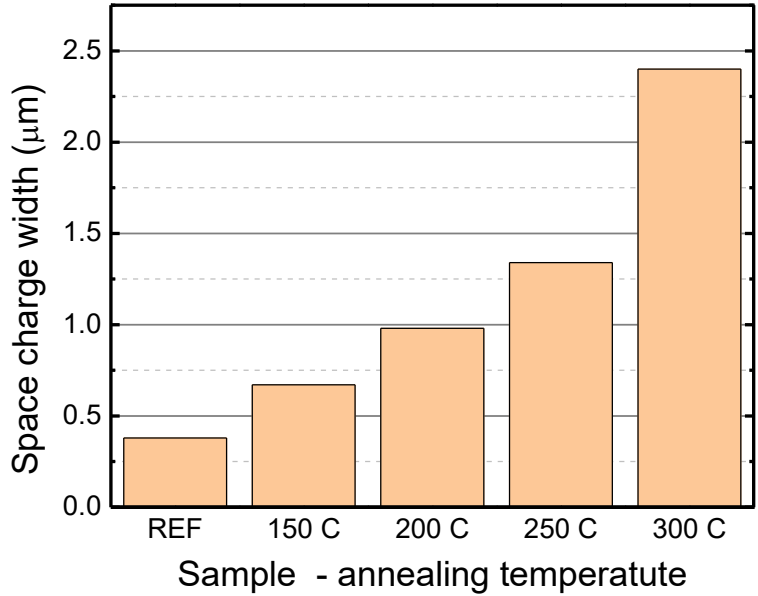

**Figure 7: Space charge region width at Room temperature derived from C-V measurements.** The calculated space charge region width increases up to the complete CIGS layer thickness of 2.4 µm when the annealing temperature increases suggesting that charges are compensated progressively as the annealing temperature increases.

It is evident that the net acceptor concentration gradually decreases with the annealing temperature. In order to verify whether the accompanying increase of the depletion layer width might explain the observed open circuit voltage losses we plotted in Fig. 8 room temperature $V_{OC}$ vs. free hole concentrations. If Cd diffusion into the CIGSe layer would only result in an increase of the SCR width, then, assuming that $V_{OC}$ is limited by SCR recombination, it should depend on $p$ in the following way

$$V_{oc}(p) = V_{oc}^{ref} - \frac{1}{2}\frac{AkT}{q}\ln\left(\frac{p^{ref}}{p}\right), \qquad (2)$$

where $V_{OC}^{ref}$ and $p^{ref}$ stand for the open circuit voltage and free hole concentration of the reference sample, respectively, A is the ideality factor, k – the Boltzmann constant, and T – the temperature of the measurement. Equation (2) is represented in Figure 8 by the dashed red line. It is plain to see that the slight $V_{OC}$ decrease for annealing done at 150° C and 200° C can be ascribed to this mechanism. This is equivalent to a gradual compensation of the charge carriers doping within the CIGSe material as the annealing temperature increases in agreement with Kijima et al. [64]. Thus, the gradual compensation of charge carriers in the absorber layer goes along with the observed chemical changes inside the absorber layer; namely with the progressive Cd in-diffusion. More precisely, Cd in diffusion leads to the formation of $Cd_{Cu}^+$ donor like defects by compensating progressively the $V_{Cu}^-$ vacancies available in the CIGSe layer. However, annealing at higher temperatures results in a drastic change in the recombination mechanism limiting $V_{OC}$ as the experimental data severely deviate from this simple model. One could explain this change by an assumption that at higher temperatures Cd diffusion not only decreases the free hole concentration but also introduces a large density of recombination centers. Although we cannot exclude this possibility, the fact that the extrapolated $qV_{OC}$ is much smaller than the absorber bandgap indicates that the recombination mechanism responsible for $V_{OC}$ losses is rather related to interface recombination. Interestingly, samples which suffer from a large $V_{OC}$ deficit, i.e. those annealed at 250° C and 300° C exhibit also a huge cross-over effect (not shown). This seems to corroborate our conclusion as the cross over effect in CIGSe solar cells is usually ascribed to the presence of a photosensitive secondary barrier in the interface region [65].

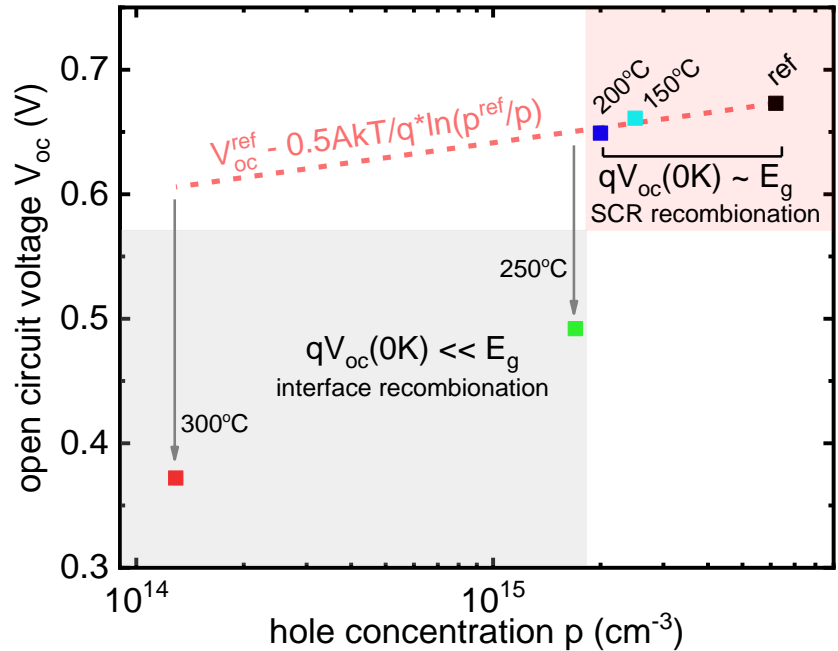

**Figure 8: Room temperature open circuit voltage $V_{oc}$ vs. free hole concentration p after light soaking**. The dashed line represents the dependence of $V_{oc}$ on free hole concentration predicted by the model (equation (2)) in which only the depletion width controls the SCR recombination, i.e. the concentration of recombination centers is constant.

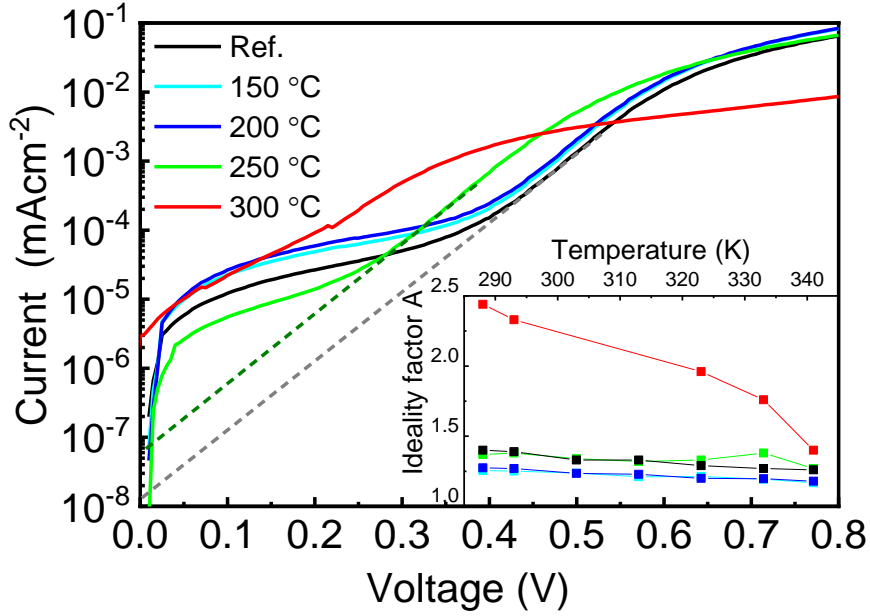

**Figure 9: Room temperature dark current – voltage characteristics and temperature dependence of ideality factor *A* in dependence of annealing temperature**. Dashed extrapolation lines of the linear part of I-V curves indicate that the ideality factor for the 250°C sample is nearly the same as for the Reference sample and that the main difference lies in the saturation current.

Figure 9 depicts the room temperature dark log(I)-V characteristics with the temperature dependence of the ideality factors A(T) in the inset in dependence of the annealing temperature. As one can see, for all samples except that annealed at 300° C, the ideality factors are in the range of 1.2 – 1.4 and they only slightly increase with decreasing temperature. This is easily understood for samples in which the transport is limited by SCR recombination, i.e. Ref., 150° C and 200° C. On the contrary, for the sample annealed at 300° C the ideality factor exhibits much stronger temperature dependence with the values of A>2 which indicates the influence of tunneling enhanced interface recombination. Such situation would require a very non-uniform defect distribution, i.e much larger concentration of defects in a thin close-to-interface CIGS layer in order to assure large enough local electrical field in an overall highly compensated sample. Interestingly, the existence of such layer was postulated to explain the cross-over effect [65]. The most puzzling is the sample annealed at 250°C which seems to exhibit features of both types discussed above: on the one hand $V_{OC}$ is limited by the interface recombination ($E_a<E_g$), but on the other hand its ideality factors behave as if the transport were influenced mainly by SCR. We suppose that both recombination paths coexist here, and their impact on transport properties strongly depends on illumination and/or voltage bias. This suggests that annealing of CIGSe devices above 200° C modifies also defects in the interface region. However, in order to clarify the detailed mechanism would require much larger set of samples covering the temperature range between 200° C and 300° C.

## 4  Conclusions

In the present work we have used complementary microanalysis techniques, including APT, temperature dependent J-V measurements, EQE, as well as C-V measurements to investigate the Cd diffusion inside the absorber layer and its consequence on the cell performance. The temperature dependent J-V measurements have shown progressive decrease in the cell's efficiency with the applied heat treatment from 150°C to 300°C and a change of predominant recombination mechanism from bulk to interface recombination above 200°C of annealing.

Using APT, we have obtained the chemical composition across the CdS/CIGSe interface independently of its high roughness and with sub-nanometer resolution. These results have shown, that the CdS/CIGSe interface is not sharp but consists of an interfacial zone (2 - 6.5 nm wide) where a gradient of constituent elements belonging to CdS and CIGSe appears. An enhanced Cd in-diffusion inside the CIGSe was observed whenever a low GGI ratio ($\leq 0.3$) occurred at the interface. This indicates the presence of $V_{Ga}$ as a channeling defect for Cd in-diffusion inside the near surface CIGSe layer. A Ga and Cu depletion was found to enhance the inversion at the CdS/CIGSe which reduces recombination of charge carriers. Hence, a too high Ga content at the CIGSe surface prevents the formation of type inversion at the interface. This indicates two main requirements that need to be fulfilled during growth: i) limited composition range for Ga at the CIGSe surface and ii) controlled composition of each individual grain.

In addition, using APT experiments we have studied the Cd diffusion into the CIGSe layer at subsequent temperatures of annealing. Considerable amounts of Cd were found inside the CIGSe layer as the annealing temperature increases that could be correlated with the decrease in the cell efficiency. Further investigations of element redistribution inside the CIGSe layer combined with C-V measurements have supported the formation of $Cd_{Cu}^{+}$ donor like defects deep inside the *p*-type CIGSe which lead to a strong compensation of the CIGSe layer. At temperatures higher than 200°C interface recombination further leads to a to strong deterioration of cell performance.

**Acknowledgement**

The authors would like to thank Prof. Dr. hab. Malgorzata Igalson, for the opportunity of performing electrical characterization on the solar cells investigated in this paper at the Faculty of Physics at Warsaw University of Technology. We would like to acknowledge the Deutsche Forschungsgemeinschaft for financing the fabrication of the solar cells used in this paper and the lab team at the Zentrum für Sonnenenergie- und WasserstoffForschung Baden-Württemberg in Stuttgart for the cells fabrication. We are grateful to U. Tezins and A. Sturm for their support to the FIB and APT facilities at MPIE. The authors acknowledge the International Max Planck Research School for Surface and Interface Engineering in Advanced Materials (IMPRS-SurMat) for supporting the present work. In Poland studies were funded by FOTECH-1 project granted by Warsaw University of Technology under the program Excellence Initiative: Research University (ID-UB).

# Supplementary information
**Effect of Cd diffusion on the electrical properties of the Cu(In,Ga)Se$_2$ thin-film solar cell**

Anna Koprek[a], Pawel Zabierowski[b], Marek Pawlowski[b], Luv Sharma[e], Christoph Freysoldt[a], Baptiste Gault[a,d], Roland Wuerz[c], Oana Cojocaru-Mirédin[a,f,*]

[a] Max-Planck-Institut für Eisenforschung GmbH, Max-Planck-Straße 1, 40237, Düsseldorf, Germany

[b] Faculty of Physics, Warsaw University of Technology, Koszykowa 75, 00-662 Warsaw, Poland

[c] Zentrum für Sonnenenergie- und Wasserstoff-Forschung Baden-Württemberg, Meitnerstrasse 1, 70563 Stuttgart, Germany

[d] Department of Materials, Royal School of Mines, Imperial College, Prince Consort Road, London, SW7 2BP, UK

[e] Department of Mechanical Engineering, Technische Universiteit Eindhoven

[f] I. Institute of Physics, RWTH University Aachen, Sommerfeldstraße 14, 52074 Aachen, Germany

**\*Corresponding Authors:** cojocaru-miredin@physik.rwth-aachen.de;

**APT data reconstruction approach:**

The APT data collected allow to create a three-dimensional reconstruction. Here, using the commercial software package IVAS 3.6.10 (Cameca Instruments Inc.), we explored different reconstruction parameters, i.e. image compression factor (ICF) and field reduction factor $k_f$ with the goal to optimize the resulting reconstruction. The reconstructions were built using the so-called "shank angle" protocol [66] that gives the evolution of the radius of curvature of the specimen as a function of the analyzed depth. We also used the model without assuming tangential continuity, which imposes similar radii of the spherical apex and underlying truncated cone.

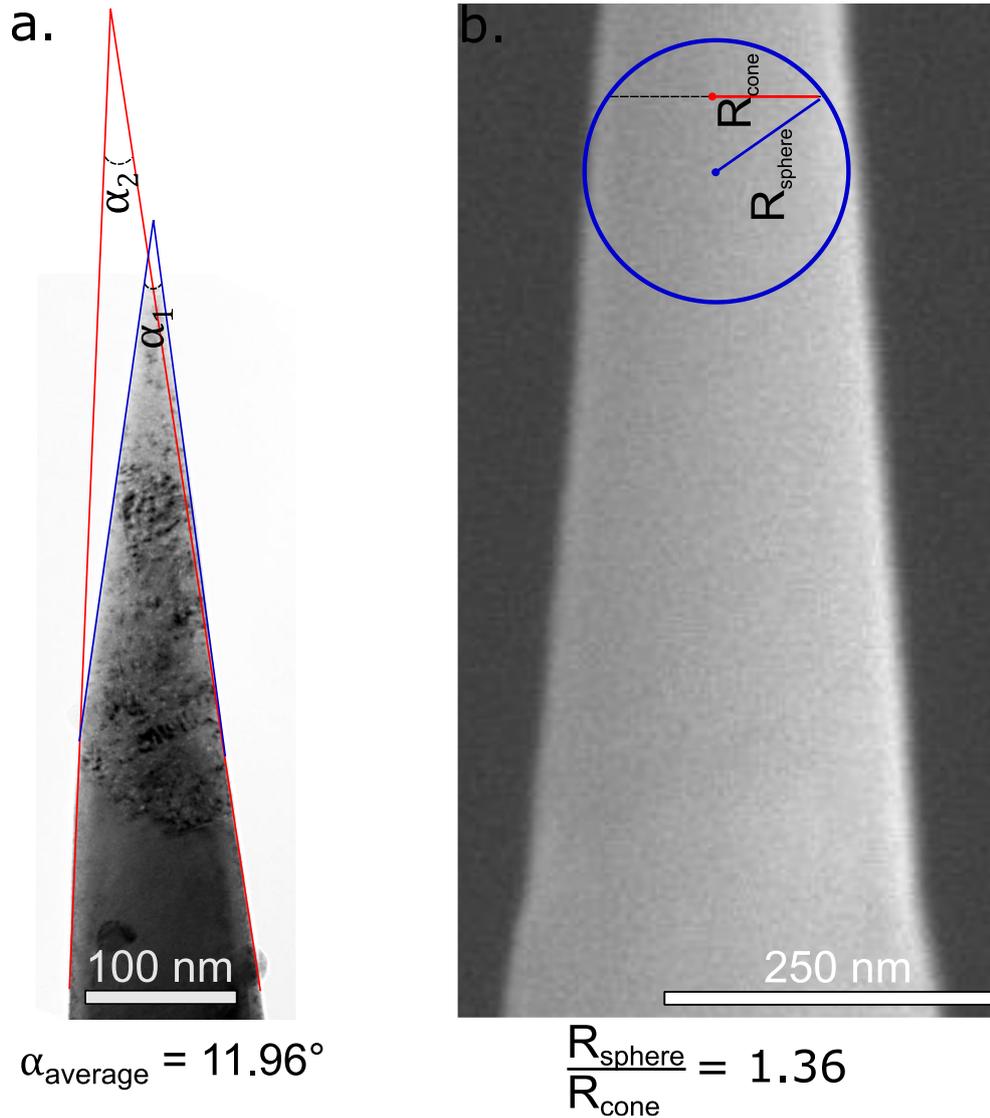

**Figure S1:** (a) TEM micrograph used to measure an averaged shank angle of the APT specimen (before performing APT investigations). (b) SEM image taken after the APT experiment that is used to estimate sphere-to-cone radius ratio $R_{SC}$.

Based on the ratio of the ionization charge states of the same element (here $Cu^{2+}/Cu^{+}$), the average evaporation field over the course of an experiment was determined to be approx. 22 $Vnm^{-1}$, based on the work of Kingham[67]. This model enables to estimate the values of the electrostatic field at the specimen's surface, which are more representative than the values typically derived at 0 K from Tsong's model[68]. The geometric parameter needed for the reconstruction i.e. averaged shank angle $\alpha_{average}$ and sphere-to-cone radius ratio $R_{SC}$, were measured from the SEM (or TEM) images that were taken respectively before and after the APT experiment. Within each APT reconstruction, a Cd iso-composition surface was created in order to visualize the 3D morphology of the CdS/CIGSe interface. The threshold was selected at point where the S and Se concentration profiles intersect, as very little or no diffusion is expected to occur for these elements[9]. The obtained Cd iso-composition surface was progressively rotated around the {z}-axis and compared with the contour and inclination angle of the projected CdS/CIGSe interface obtained by TEM. Among three examined specimens best match between projected and reconstructed CdS/CIGSe interface was found for ICF=1.4 and $k_f$=4.

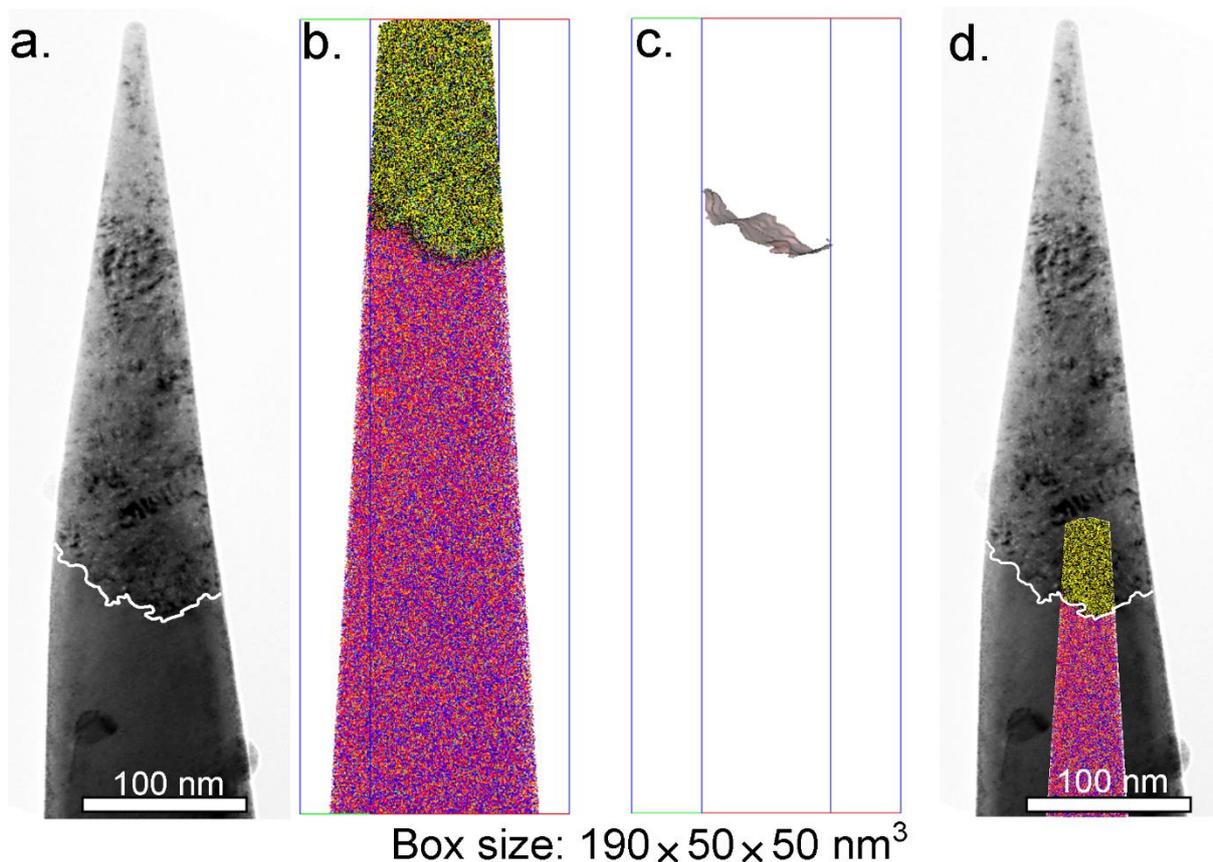

**Figure S2:** (a) TEM micrograph presenting projection of the CdS/CIGSe interface - white contour. (b) image of the APT reconstruction obtained using ICF=1.4; $k_f$=4. (c) reconstructed CdS/CIGSe interface shown as Cd iso-composition surface with a threshold of 32 at.%. (d) overlap between reconstructed and projected interface. For clarity, here the tip-shape specimen was further sharpened by FIB-SEM to reduce the thickness of the CdS buffer layer and then investigated by APT. This explain the big difference in size between the TEM imaging and APT reconstruction.

In addition, the analysis of the geometric parameters of all investigated here APT specimens showed an almost constant sphere-to-cone radius ratio of $R_{CS}$=1.25. Likewise, limited variations in the averaged shank angle from 10–20° were observed. This was also corroborated by the rather small range of fluctuations of the mass resolution at full-width half maximum (FWHM) of 566±25 Da (Da stands for Dalton) for the peak located at 63 Da identified as $^{63}Cu^+$. This indicates that the cooling of the APT specimen after every laser pulse was sufficiently fast. Hence, one can expect that the shape of the specimen, that defines the field distribution around the tip, will almost exclusively be driven by the evaporative behavior of specimen constituents. The values of the reconstruction parameters obtained for three different APT specimens were very consistent. In addition, the APT specimen's geometrical parameters were very similar, and all experiments were performed on the same microscope with the same experimental parameters (base-temperature of 60 K, laser energy of 0.1 nJ and repetition rate of 100 kHz). It hence seemed reasonable to assume that the same reconstruction parameters, ICF=1.4 and $k_f$=4, could be used across all datasets reported in this paper.

**Deoverlapping peaks in the APT mass spectrum:**

Before presenting composition profiles obtained from APT reconstructions, an extensive analysis of the APT mass spectrum has to be performed, which includes elemental identification of each peak, background subtraction, and separation of the overlaps of ions with the same mass-to-charge ratio that appear in the mass spectrum. The analysis of the rough CdS/CIGSe interface was hence conducted within cylinders (20 nm-diameter and of 60 nm-long) cut out from the APT reconstructions perpendicularly to a flat region of the CdS/CIGSe interface as shown in Fig. S3. In this way, the influence of the roughness of the CdS/CIGSe interface on our results is expected to be limited. The long range diffusion was investigated using a series of 20nm-wide slices cut out from the reconstruction along the direction defined by this cylinder, thereby allowing direct correlation of the interface composition with the long range concentration profiles (See Fig.4). The width of the slices was systematically varied from 5 to 20 nm without noticeable differences.

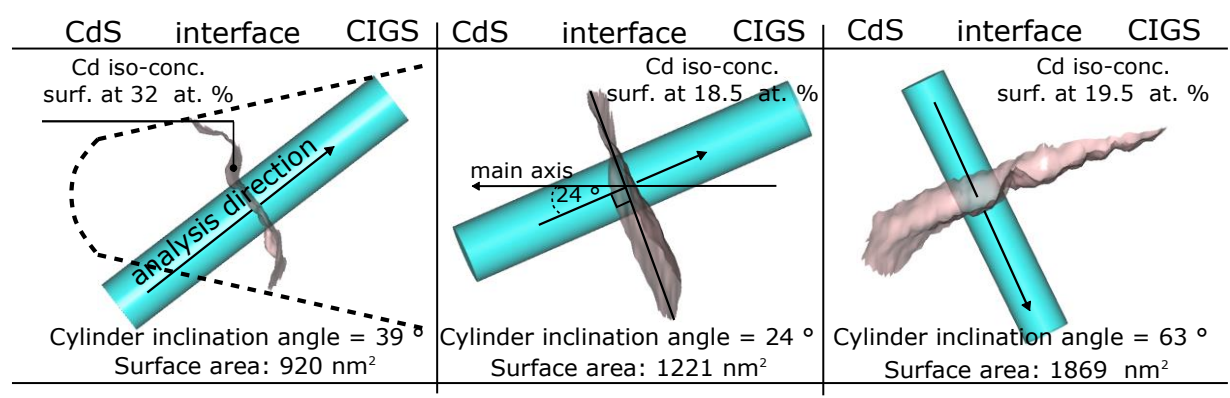

**Figure S3:** Three examples highlighting the complexity of the CdS/CIGSe interface represented as an Cd iso-composition surface (see text for details). The turquoise cylinders are positioned perpendicularly to flat regions of the iso-concentration surface and are used to export sub-volume of the APT reconstruction for chemical analysis of the interface.

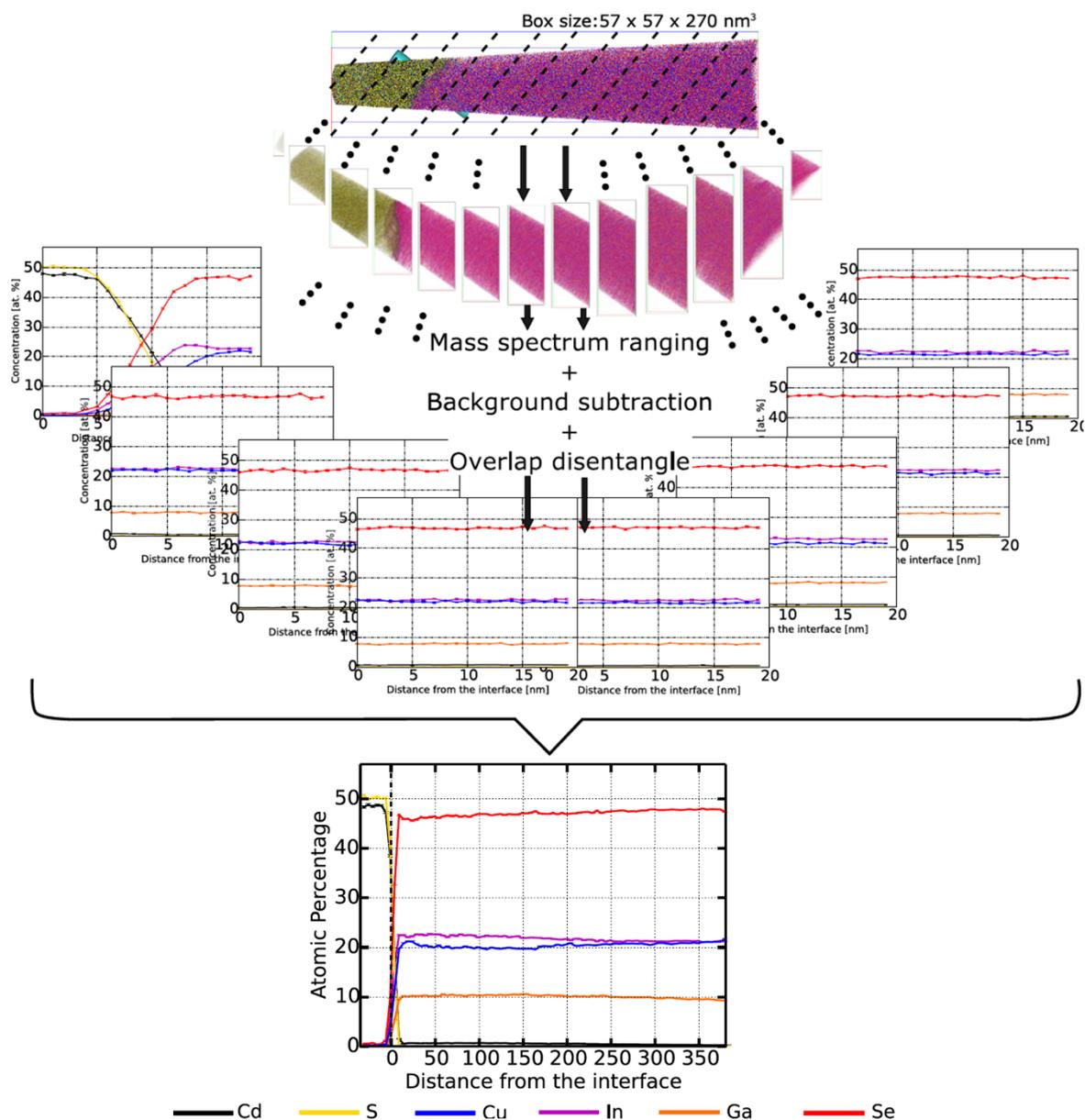

**Figure S4:** Long range diffusion profiles obtained from dividing the APT reconstruction into 20 nm-wide slices (see text for details). Within each slice, extended mass spectrum analysis were performed, and a single compositional profile was then built from the combination of each slice.

Once region-of-interest is exported, the local mass spectrum is calculated. For each peak, the lower and higher bound of a specific range of mass-to-charge ratios comprising this peak are defined and assigned an identity as a specific atomic or molecular ion. Here, each peak is automatically ranged based on its full-with-at-tenth maximum (FWTM), which is expected to give a rather accurate composition[69]. Each peak is influenced by the background in the spectrum that partly originates from e.g. the dark current of the detector that forms a random component, but also the tails from peaks at lower mass-to-charge ratios originating from thermalization of the specimen in laser assisted APT or energy deficit in voltage-pulsed APT. The background and peak overlap influence the capacity of the technique to spatially resolve the constituent elements, and potentially leading to misinterpretation of the analysis, as presented in Fig. S5.

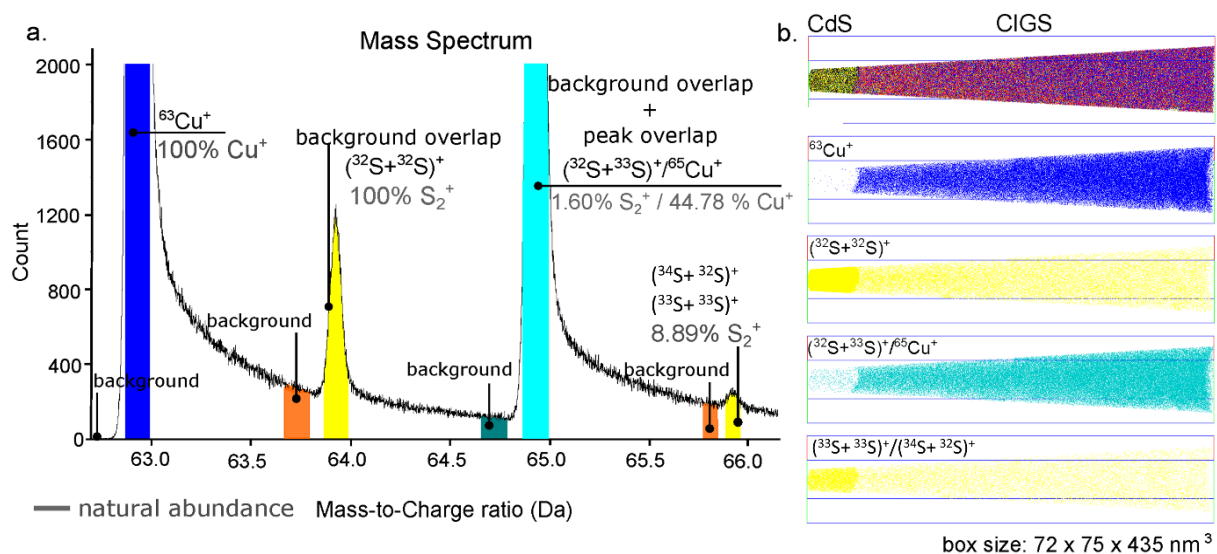

**Figure S5:** (a) APT mass spectrum showing peaks and ranges (in color) identified as Cu and S isotopes that overlap at 65 Da and height of the local background coming from thermal tails of the identified ions that overlap on the adjusted peak located on the side of higher mass-to-charge ratio. (b) illustration of the background and peak overlaps shown in a 3D elemental map.

To overcome this issue, for each peak, an additional range is defined to quantify the local level of background, and help with reducing the influence of the background on our reported concentration values. This additional range is created, with the same width as the corresponding range, shifted with respect to the position of the peak by 1.5 FWTM. The width and position of the background range were defined to avoid subtraction of actual signal from the peak and simultaneously to avoid overlap with possible neighboring peaks, as shown in Fig.S6 (a).

One-dimensional concentration profile was hence obtained for each of the defined ion and molecular ion on either side of the CdS/CIGSe interface (see Fig. S6 (a)). The background was then subtracted, leaving only statistical fluctuation (See Fig. S6 (b)). These profiles were subsequently used to deoverlap ion species, which significantly affect the determination of the interface composition. First, we defined the specific isotopes of each species and their ionic states that constitutes each overlap. Second, based on the natural abundance of the isotopes belonging to the same species but not affected by an overlap, we calculated within every bin of the profile the relative number of counts contributing to the overlap. As an example of result, the one-dimensional concentration profiles for $^{65}Cu^+$ and $^{32+33}S_2^+$ ions is shown in in Fig. S6.

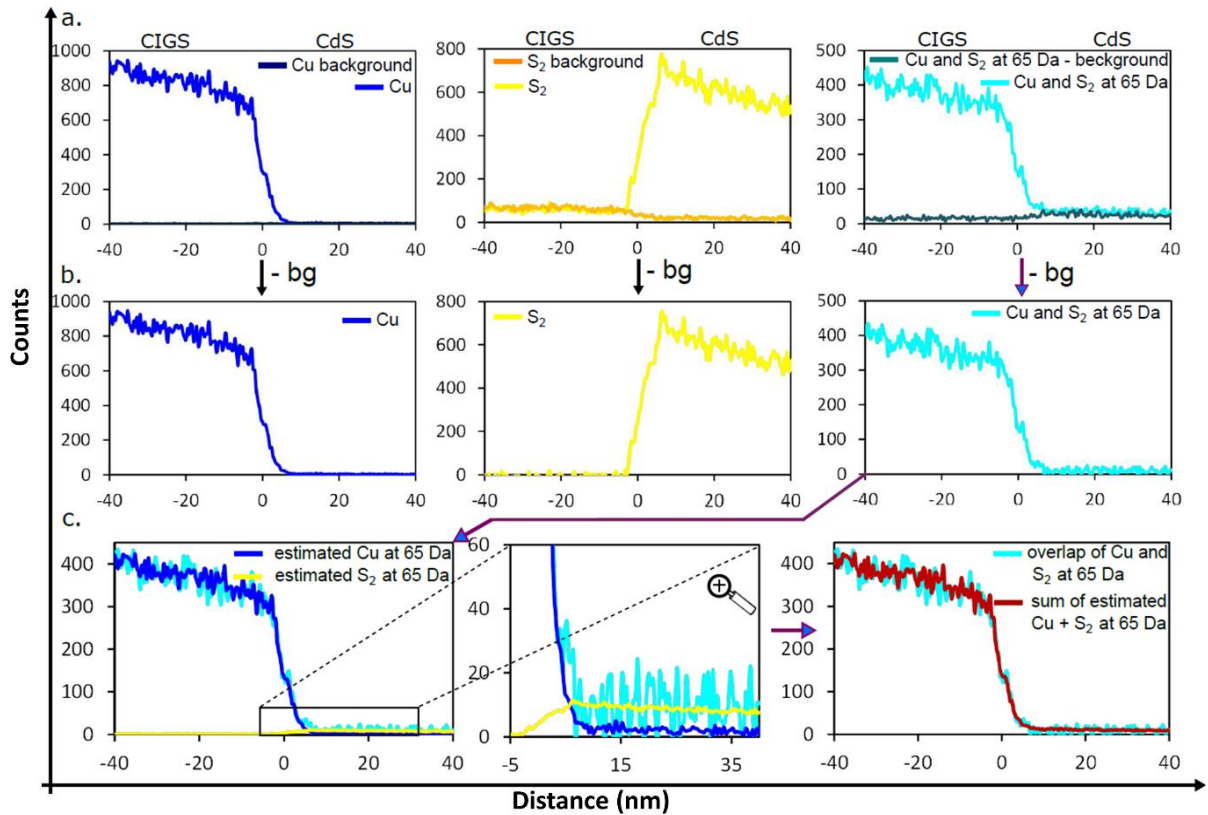

**Figure S6:** Deoverlapping of the overlapped peaks at 65 Da. Based on the 1D distribution, intensity and natural abundance of $Cu^+$ and $S_2^+$ isotopes that are not influenced by any overlap the distribution of $^{65}Cu^+$ (dark blue) and $^{32+33}S_2^+$ (yellow) hidden in the peak located at 65 Da could be calculated. Estimated distributions were further contrasted with the distribution of the signal at 65 Da (turquoise). A close-up on the interface region reveals separation of the estimated $^{65}Cu^+$ and $^{32+33}S_2^+$ signals at the phase boundary (a.). 1D concentration profile of the signal located at 65 Da (turquoise) and sum of the estimated signals for $^{65}Cu^+ + ^{32+33}S_2^+$ (red) (b.).

The magnified image of the interface region shows that the applied procedure successfully separates signals belonging to different phases. In order to find out whether the calculated number of counts per each species in each distance reliably represents the counts inside the overlapped peaks, the estimated distribution of the isotopes of the constituents were added together and compared with the measured distribution as shown in Fig. S6c. This procedure was applied to deoverlapping peaks for each of the overlaps identified in the mass spectrum (such as Cu and S overlap at 32.5 Da and 65 Da, Cd and In overlap at 56.5 and 113 Da, and group of overlaps of CuSe and CdS between 136 and 147 Da). This allowed to eliminate artifacts in the composition profiles that lead to misinterpret overlaps as interdiffusion of native elements across the CIGSe/CdS interface (see Fig. 4, main text). It should be mentioned that the applied method was developed by the authors of this paper for the purpose of interface and long range diffusion profile analysis in the CdS/CIGSe system. The subsequent steps of this method were automatized within a python script.